\documentclass[%
 reprint,
superscriptaddress,
 amsmath,amssymb,
aps,
prb,
]{revtex4-1}

\usepackage{graphicx}% Include figure files
\usepackage{dcolumn}% Align table columns on decimal point
\usepackage{bm}% bold math
\usepackage{hyperref}% add hypertext capabilities
\usepackage[mathlines]{lineno}% Enable numbering of text and display math
\usepackage{color,soul}
\usepackage[utf8]{inputenc}

\newcommand{\red}[1]{{\color{red}#1}}

\begin{document}
\setstcolor{red}
\setul{0.3ex}{0.3ex}
\preprint{APS/123-QED}

\title{Emergence of glassy features in halomethane crystals}% Force line breaks with \\
%thanks{A footnote to the article title}%

\author{Manuel Moratalla}
\affiliation{Laboratorio de Bajas Temperaturas, Departamento de Física de la Materia Condensada, Condensed Matter Physics Center (IFIMAC) and Instituto Nicolás Cabrera, Universidad Autónoma de Madrid, Francisco Tomás y Valiente 7, 28049 Madrid, Spain}
\author{Jonathan F. Gebbia}
\affiliation{Grup de Caracterizació de Materials, Departament de Fisica, EEBE and Barcelona Research Center in Multiscale Science and Engineering, Universitat Politècnica de Catalunya, Eduard Maristany, 10-14, 08019 Barcelona, Catalonia}
\author{Miguel Angel Ramos}
\affiliation{Laboratorio de Bajas Temperaturas, Departamento de Física de la Materia Condensada, Condensed Matter Physics Center (IFIMAC) and Instituto Nicolás Cabrera, Universidad Autónoma de Madrid, Francisco Tomás y Valiente 7, 28049 Madrid, Spain}
\author{Luis Carlos Pardo}
\affiliation{Grup de Caracterizació de Materials, Departament de Fisica, EEBE and Barcelona Research Center in Multiscale Science and Engineering, Universitat Politècnica de Catalunya, Eduard Maristany, 10-14, 08019 Barcelona, Catalonia}
\author{Sanghamitra Mukhopadhyay}
\affiliation{ISIS Facility, Rutherford Appleton Laboratory, Chilton, Didcot, Oxfordshire OX11 0QX, United Kingdom}
\author{Svemir Rudi{\'c}}
\affiliation{ISIS Facility, Rutherford Appleton Laboratory, Chilton, Didcot, Oxfordshire OX11 0QX, United Kingdom}
\author{Felix Fernandez-Alonso}
\affiliation{ISIS Facility, Rutherford Appleton Laboratory, Chilton, Didcot, Oxfordshire OX11 0QX, United Kingdom}
\affiliation{Department of Physics and Astronomy, University College London, Gower Street, London WC1E 6BT, United Kingdom}
\author{Francisco Javier Bermejo}
\affiliation{Instituto de Estructura de la Materia, Consejo Superior de Investigaciones Científicas, CSIC, Serrano 123, 28006 Madrid, Spain}
\author{Josep Lluis Tamarit}
\email{josep.lluis.tamarit@upc.edu}
\affiliation{Grup de Caracterizació de Materials, Departament de Fisica, EEBE and Barcelona Research Center in Multiscale Science and Engineering, Universitat Politècnica de Catalunya, Eduard Maristany, 10-14, 08019 Barcelona, Catalonia}

%\author{Manuel Moratalla}
% \altaffiliation[Also at ]{Physics Department, XYZ University.}%Lines break automatically or can be forced with \\
%\author{Second Author}%
% \email{Second.Author@institution.edu}
%\affiliation{%
% Authors' institution and/or address\\
% This line break forced with \textbackslash\textbackslash
%}%

%\collaboration{MUSO Collaboration}%\noaffiliation

%\author{Charlie Author}
% \homepage{http://www.Second.institution.edu/~Charlie.Author}
%\affiliation{
% Second institution and/or address\\
% This line break forced% with \\
%}%
%\affiliation{
% Third institution, the second for Charlie Author
%}%
%\author{Delta Author}
%\affiliation{%
% Authors' institution and/or address\\
% This line break forced with \textbackslash\textbackslash
%}%

%\collaboration{CLEO Collaboration}%\noaffiliation

\date{\today}% It is always \today, today,
             %  but any date may be explicitly specified

\begin{abstract}
Both structural glasses and disordered crystals are known to exhibit \textit{anomalous} thermal, vibrational and acoustic properties at low temperatures or low energies, what is still a matter of lively debate. To shed light on this issue, we have studied the halomethane family CBr$_{n}$Cl$_{4-n}$ ($n = 0,1,2$) at low temperature where, despite being perfectly translationally-ordered stable monoclinic crystals,  \textit{glassy} dynamical features had been reported from experiments and molecular dynamics simulations. For $n = 1,2$ dynamic disorder originates by the random occupancy of the same lattice sites by either Cl or Br atoms, but not for the ideal reference case of CCl$_4$. Measurements of the low-temperature specific heat ($C_p$) for all these materials are here reported, which provide evidence of the presence of a broad peak in Debye-reduced $C_{p}(T)/T^{3}$ and in the reduced density of states ($g(\omega)/\omega^2$) determined by means of neutron spectroscopy, as well as a linear term in $C_{p}$ usually ascribed in glasses to two-level systems in addition to the cubic term expected for a fully-ordered crystal. Being CCl$_{4}$ a fully-ordered crystal, we have also performed density functional theory (DFT) calculations, which provide unprecedented detailed information about the microscopic nature of vibrations responsible for that broad peak, much alike the “boson peak” of glasses, finding it to essentially arise from a piling up (at around $3-4$ meV) of low-energy optical modes together with acoustic modes near the Brillouin-zone limits.
%\begin{description}
%\item[Usage]
%Secondary publications and information retrieval purposes.
%\item[PACS numbers]
%May be entered using the \verb+\pacs{#1}+ command.
%\item[Structure]
%You may use the \texttt{description} environment to structure your abstract;
%use the optional argument of the \verb+\item+ command to give the category of each item. 
%\end{description}
\end{abstract}

\pacs{Valid PACS appear here}% PACS, the Physics and Astronomy
                             % Classification Scheme.
%\keywords{Suggested keywords}%Use showkeys class option if keyword
                              %display desired
\maketitle

%\tableofcontents

\section{\label{sec:Intro}Introduction}

Structural (conventional) glasses (SG) are amorphous solids \cite{ElliottSR1990} lacking both translational and orientational long-range order, usually obtained by quenching or freezing the supercooled liquid \cite{debenedetti2001supercooled, cavagna2009supercooled, ediger2012perspective}. \red{Orientational glasses (OG) consist of crystals with positional order and orientational disorder and exhibit similar glassy properties that conventional glasses\cite{Sherwood1978, binder2005glassy}. Both kinds of glasses} share a complex atomic dynamics that cannot be described by the standard wave-like oscillations (phonons) observed in crystalline solids. Such complexity modifies the low-frequency excitations (the low-frequency vibrational density of states, VDoS, $g(\omega)$) and thus results on thermodynamic and transport anomalies with respect to their well-known crystalline counterparts.

	The leading macroscopic property characterizing those thermal anomalies in glasses is the specific heat ($C_p$). In the usual Debye-reduced representation in terms of $C_{p}(T)/T^{3}$ as a function of $T$, a hump appears with a maximum at low temperatures (typically between 4 and 12 K) deviating from the horizontal line predicted by the Debye theory. There is nowadays a broad consensus that this heat capacity excess comes from additional low-frequency phonon-like excitations \cite{buchenau1986low, Bagglioli2018}, which is evidenced in plots of $g(\omega)/\omega^2$ by a correlated broad peak at $\omega_{BP}\approx 2-4$ meV (ca. 1 THz), the so-called \textit{boson peak} (BP) \cite{Phillips1981Amorphous}. In addition, the specific heat at low temperatures also displays a characteristic property of glasses: the apparently universal existence of tunneling two-level systems (TLS). The systematic observation of a linear term at low temperature in $C_{p}$ for non-metallic glasses \cite{zeller1971thermal,Phillips1981Amorphous} was soon ascribed to the ubiquitous presence of a random distribution of TLS as proposed by the Tunneling Model \cite{phillips1972tunneling,anderson1972anomalous}. These and some other features, such as the strong contribution of TLS to thermal and acoustic properties around 1$-$2 K, or the universal plateau both in the thermal conductivity and the acoustic internal friction at about a few K, have been considered as inescapable properties of structural glasses (SG) first \cite{Phillips1981Amorphous} and, afterwards, also of orientational glasses (OG), either mixed crystals such as alkali-halide/alkali-cyanide crystals \cite{knorr1985x,de1986low,Hochli1990,watson1995tunneling} or orientationally-disordered crystals (called “glassy crystals”) obtained by quenching high lattice symmetry plastic crystals with rotational disorder \cite{bonjour1981low,ramos1997quantitative,talon2002E,vdovichenko2015thermal,vispa2017thermodynamic}. Furthermore, some of these glassy anomalies have also been reported in other disordered solids, including quasicrystals \cite{lassjaunias1997investigation,pohl2002low}, \red{some kind of incommensurate modulated crystals with broken translational periodicity} \cite{remenyi2015incommensurate}, and low-dimensional organic crystals with minimal amounts of disorder \cite{gebbia2017glassy,romanini2012emergence}. All such disparate systems share some type of frustration, namely the presence of competing interactions of energetic or entropic nature which make it extremely difficult the attainment of a ground state, thus leading the system to get trapped into long lived metastable states. Some recent works have associated the excess low-frequency modes in glassy systems with transverse-like vibrations \cite{shintani2008universal,schirmacher2006thermal,zhang2017experimental,chumakov2011equivalence} through the argument that transverse vibrations are more sensitive to the lack of order than longitudinal ones, while other \cite{ruffle2006glass} attributed the boson peak to the Ioffe-Regel limit of longitudinal phonons. Another approach is provided by the Soft-Potential Model (SPM), which postulates a coexistence of sound waves and quasilocalized modes, either tunneling TLS or soft vibrational modes \cite{karpov1983zh,buchenau1992interaction,parshin1994interactions,Ramos1998tunneling}.

Here we report on results pertaining a system composed by globule-like molecules which because of the weak and nearly spherical intermolecular interactions at play cannot easily \cite{yamamuro2003neutron,Chua2016Glass} form structurally amorphous phases but rather form rotator-phase crystals \cite{criado2000rotational} which exhibit phenomena quantitatively close to those shown by fully disordered solids. The materials under study are members of the halomethanes family, namely CBr$_{n}$Cl$_{4-n}$ with $n=0,1,2$. They present orientationally disordered (OD) face-centered cubic or rhombohedral (plastic) crystalline phases near below room temperature \cite{ohta1995heat,tamarit2008high,zuriaga2009new,zuriaga2012dynamic}. As revealed by neutron diffraction \cite{binbrek1999crystal}, on cooling from the OD phases, all of them transform into low-symmetry crystals (monoclinic C2/c, with $Z=32$ molecules per unit cell and an asymmetric unit with $Z^{\prime}=4$ molecules) below 200-220 K down to the lowest temperatures \cite{tamarit2008high}, hence devoid \textit{a priori} of any kind of translational or orientational disorder. Nevertheless, quasi-tetrahedral CBr$_{n}$Cl$_{4-n}$ molecules exhibit (for $n\neq0$) statistical “occupational disorder” between Cl and Br atoms \cite{ohta1995heat,zuriaga2012dynamic,zuriaga2009new}, which persist down to $\sim90$ K where a calorimetric transition much alike that exhibited by the canonical glass-transition signals the transition into a frozen disordered state where molecular reorientations cannot be detected with the available experimental means \cite{ohta1995heat}.

The dynamics of such disorder in halomethane crystals has been studied by means of dielectric spectroscopy (for $n\neq0$) and nuclear quadrupole resonance (NQR) spectroscopy (for $n=0$) \cite{zuriaga2009new,zuriaga2012dynamic}. Both techniques and the help of extensive molecular dynamics simulations have demonstrated the existence of large-angle rotations of tetrahedra about their higher molecular symmetry axes (CBrCl$_{3}$, and CBr$_{2}$Cl$_{2}$ with C$_{3v}$ and C$_{2v}$ point-group symmetries, respectively) due to the statistical occupancy of 75\% for Cl and 25\% for Br atoms in the case of CBrCl$_{3}$ \cite{parat2005polymorphism,caballero2017dynamic}, and 50\% for Cl and 50\% for Br atoms in CBr$_2$Cl$_2$ \cite{barrio2008polymorphism}. As for CCl$_4$ (T$_d$ point-group symmetry), results have revealed identical dynamics of the monoclinic phase as a function of temperature \cite{zuriaga2009new,zuriaga2011rotational} and, consequently, the four nonequivalent molecules of the asymmetric unit ($Z^{\prime}=4$) perform reorientational jumps each one with slightly different time scales due to their different crystalline (and well-defined) environment. Moreover, it was found that the molecules sitting on nonequivalent crystal sites exhibit dynamics characterized by different timescales. Even more surprising, the results from dielectric spectroscopy revealed relaxation patterns for the Bromine-Chlorine-Methane crystals astonishingly close to those exhibited by glass-forming materials near the glass transition. Therefore, these structurally-ordered crystals seem to present dynamical heterogeneity \cite{richert2002heterogeneous,berthier2011dynamical,caballero2017dynamic,zuriaga2012dynamic,caballero2016dynamic,zuriaga2011rotational}.

 Although the occupational disorder should be absent in the reference case of tetrachloromethane, CCl$_4$ \cite{hicks1944heat,atake1971heat,Bagatskii1971}, the same dynamics has been strikingly revealed by NQR measurements \cite{zuriaga2009new} and molecular dynamics simulation \cite{caballero2016dynamic}. On the other hand, thermal conductivity measurements \cite{krivchikov2015effects} in the same three halomethane crystals have shown an absence of the characteristic glassy anomalies at moderately low temperatures. The thermal conductivity of these molecular crystals was found to display two distinct temperature regions, the boundary of which roughly coincided with the abovementioned glass-like transitions. The absence of the ubiquitous plateau in the thermal conductivity of glassy systems was interpreted as a consequence of a large mismatch between frequencies characteristic of heat-carrying phonons and those expected for reorientational jumps.
    The aim of this work is thus to investigate whether a thermodynamic correlate of the dynamic anomalies already reported on could be found in the form of universal glassy anomalies at low temperatures/frequencies in the halomethane compounds CBr$_n$Cl$_{4-n}$ ($n = 0,1,2$) and, in particular, for the reference case $n=0$, which shows the same dynamical properties but without exhibiting crystallographic occupational disorder. Then, measurements of the specific heat $C_p$ at low temperatures have been carried out as well as measurements of the frequency distribution $g(\omega)$ or vibrational density of states (DoS) amenable to experiment. In addition, and to provide us with a reference state onto which the experimental data can be referred to, fully atomistic calculations for a fully ordered monoclinic CCl$_4$ crystal structure have been carried out by means of Density Functional Theory (DFT). Such results will enable us not only to compare the experimental $g(\omega)$ to that resulting from calculation but also to assign the most salient features of the observed distribution to well defined physical entities present in the ordered ground state. In fact, the crystalline nature of the reference monoclinic phase of CCl$_4$ enables us to monitor the behavior of the crystal lattice phonons, which, contrary to the case of amorphous materials are here well defined entities all along the Brillouin zone \cite{zuriaga2011rotational}. We will show that glass-like features such as a broad peak at around 7$-$9 K in $C_{p}/T^3$ and around 3$-$4 meV in $g(\omega)/\omega^2$ (similar to the BP of glasses), as well as a linear contribution to the specific heat at lower temperatures associated with tunneling two-level systems, are in fact observed, all of them larger in CBr$_2$Cl$_2$ than in CBrCl$_3$. Most strikingly, we have also observed such similar features in the specific heat and in the density of states of crystalline CCl$_4$. As we will demonstrate, this unexpected and provocative experimental finding cannot be attributed to a secondary source of occupational disorder brought about by the random natural isotopic distribution of 75\% ${}^{35}$Cl atoms and 25\% ${}^{37}$Cl atoms. The dynamic distortion introduced by this kind of isotopic disorder is not enough to make appear the universal set of low-energy excitations typical of the glass state. By using DFT calculations for the CCl$_4$ case, we will show that the abovementioned broad peak, arises from a piling up (at around 3$-$4 meV) of low-energy optical modes together with acoustic modes near the Brillouin-zone boundaries.

\section{\label{sec:Exp}Materials and Methods}
\subsection{Sample preparation}
Samples of CCl$_4$ and CBrCl$_3$ were obtained from Across with purity better than 99\% and used without further purification. CBr$_2$Cl$_2$ was purchased from Aldrich with a purity of 95\% and was twice fractionally distilled.

\subsection{Specific-heat measurements}
The heat capacity of the studied samples was measured in a versatile calorimetric system \cite{perez2007low}, especially designed for glass-forming liquids. Firstly using liquid nitrogen as thermal sink, we were able to concurrently measure the absolute heat capacity and to characterize the phase transitions in the range 77$-$300 K (data not shown here) employing a self-developed quasiadiabatic continuous method \cite{perez2007low,perez2013low}. Once the respective monoclinic crystals were obtained at approximately 221 K ($n = 0$), 234 K ($n = 1$), and 254 K ($n = 2$), and cooled down to 77 K, liquid nitrogen was replaced by liquid helium as thermal bath in the experimental cryostat, and accurate measurements of the heat capacity were conducted between 1.8 K and 25 K by the thermal relaxation method.  Many more details about the cryogenic system employed, electronic control, thermal sensors and heating elements can be found in \cite{perez2013low}. The calorimetric cell is essentially a thin-walled vacuum-tight closed copper can where the liquid sample has been previously inserted and carefully weighed. The calorimeter set-up is formed by a copper ring, which acts as thermal sink, and a sapphire disk suspended by nylon wires from the ring. An additional copper wire (chosen as to provide thermal relaxation times of tens of seconds) acts as thermal link between the sapphire substrate and the copper-ring reservoir. The latter has a germanium thermometer and a 50 $\Omega$ resistor (heater) that allow temperature control and increasing its temperature. In the present work, the sample cell was put on a sapphire disk and attached to it with a tiny amount of Apiezon vacuum grease to ensure good thermal contact. On the sample cell, a carbon ceramic sensor thermometer (CCS A2) is installed, also attached by Apiezon vacuum grease, and another heater (a resistor chip of 1 k$\Omega$), is put on the sapphire substrate for heating the calorimetric cell at a given fixed thermal reservoir temperature. This calorimeter set-up is firmly attached to an insert, which is sunk in the cryogenic liquid within a cryostat. With both a rotary pump and a diffusion pump a vacuum of $10^{-8}$ mbar in the internal chamber is reached. Thermometers and heaters are controlled by employing a LakeShore 336 Temperature Controller, a Keithley 224 Current Source and a Keithley microvoltmeter. A home-made program is used to automatically run the measurements. The addenda of the calorimeter using an empty copper cell was independently measured and thus subtracted from the obtained heat-capacity data. In each case, small corrections due to slight differences in the carefully measured masses of Apiezon grease, low-temperature varnish or copper in the cell were made, since their specific-heat curves are known.

\subsection{Inelastic neutron scattering measurements}
Inelastic neutron scattering measurements were performed using TOSCA and MARI spectrometers at the ISIS Pulsed Neutron and Muon Source of the Rutherford Appleton Laboratory (Oxfordshire, UK).
TOSCA indirect geometry time-of-flight spectrometer \cite{PINNA201868,Pinna28JPCS,PINNA201779,Pinna15,Stewart2014} is characterized with high spectral resolution ($\Delta E/E \sim 1.25\%$) and broad spectral range ($-24 : 4000$ cm$^{-1}$). The samples ($2–3$ g) were placed in thin walled and flat aluminum cans. To reduce the impact of the Debye$–$Waller factor on the observed spectral intensity, the sample chamber was cooled to approximately 10 K by a closed cycle refrigerator (CCR) and the spectra were recorded for 6 to 12 hours. The direct geometry spectrometer MARI has continuous detector coverage from $3.5-135$ deg. The samples were loaded into an aluminum sample holder with an annular geometry. Using a top loading CCR system with temperature range of $5-600$ K the samples were cooled down to 5 K. The inelastic neutron spectra of monoclinic phases were taken at this temperature with an incident energy of 18 meV, selected using a Fermi chopper system with a Gd foil chopper pack rotating at 200 Hz. The chosen configuration of the instrument ensured an elastic line resolution of $3\%$ $\delta E/E_i$ and a $|Q|$ range $\approx 0.4 - 6$ $\AA^{-1}$.

\subsection{DFT calculations}

Calculations of the generalized frequency spectrum $g(\omega)$, also known as vibrational density of states (VDoS), have been carried out for the reference monoclinic phase of CCl$_4$ using the CASTEP DFT code \cite{CASTEP} for a crystal lattice structure comprising one half of the 32 molecules in the unit cell. The contribution of isotopic effects on the chlorine atoms was evaluated and found to be of scant relevance \red{(see Supplemental Material at [URL will be inserted by publisher])}. Results here reported on correspond to an isotopic mixture of ${}^{35}$Cl and ${}^{37}$Cl corresponding to the natural abundance of chlorine (75\% and 25\%, respectively). Different exchange-correlation functionals \cite{DFT-KS,SEDC-G06} have been tested by calculation of the frequency spectrum at the crystal $\Gamma$-point and compared with spectra measured by neutron techniques as described below. A GGA functional supplemented with a dispersion correction PBE-G06 was found to provide the best results. The calculation of the full set of dispersion branches has been carried out spanning the full Brillouin zone \red{(see Supplemental Material at [URL will be inserted by publisher] for DFT phonon calculation details)\cite{MPgrid,RMP-Payne,CASTEP_DFPT,FOURIER_PHONON}}.
%See more details in \textit{Supp Mat}).

\section{\label{sec:RD}Results and Discussion}

The heat capacity of the different studied substances was measured below room temperature employing a calorimetric system especially designed for liquid samples at room temperature that solidify \textit{in situ} whereas being cooled and measured. The expected  liquid $\rightarrow$ plastic crystal(s) $\rightarrow$ monoclinic crystal transitions in CBr$_n$Cl$_{4-n}$ ($n = 0,1,2$)  \cite{tamarit2008high,zuriaga2012dynamic,ohta1995heat,hicks1944heat} and reverse when increasing temperature were observed and monitored by means of quasiadiabatic continuous calorimetry. Then, measurements of the heat capacity were performed between 1.8 K and 25 K by the thermal relaxation method.

\begin{figure}[h!]
	\centering
    %Cp/T^3 vs T 
		 \includegraphics[width = 0.45\textwidth , clip = true, trim = 0cm 0cm 0cm 0cm]{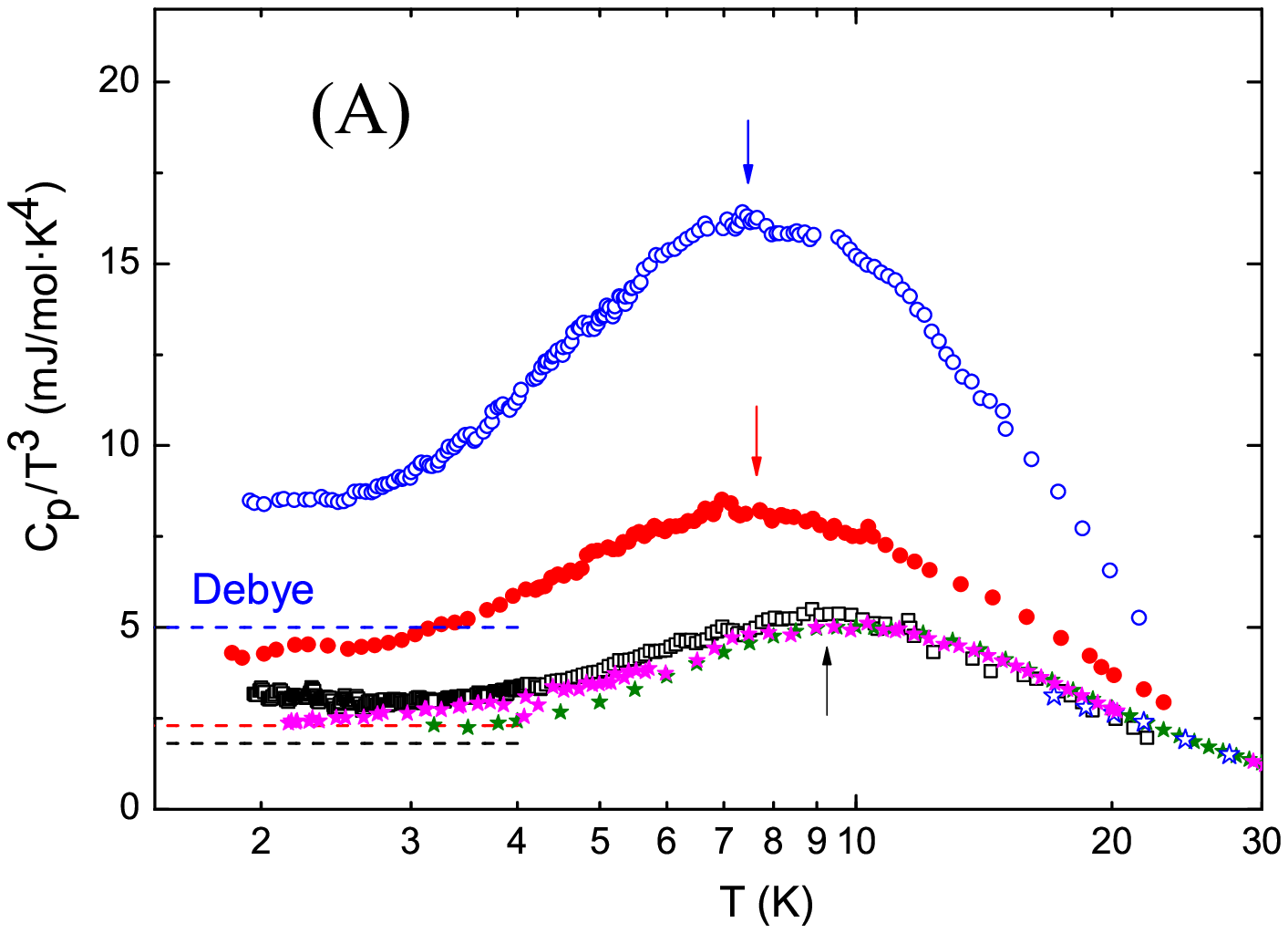}
    %Cp/T vs T^2
		 \includegraphics[width = 0.47\textwidth , clip = true, trim = 0cm 0cm 0cm 0cm]{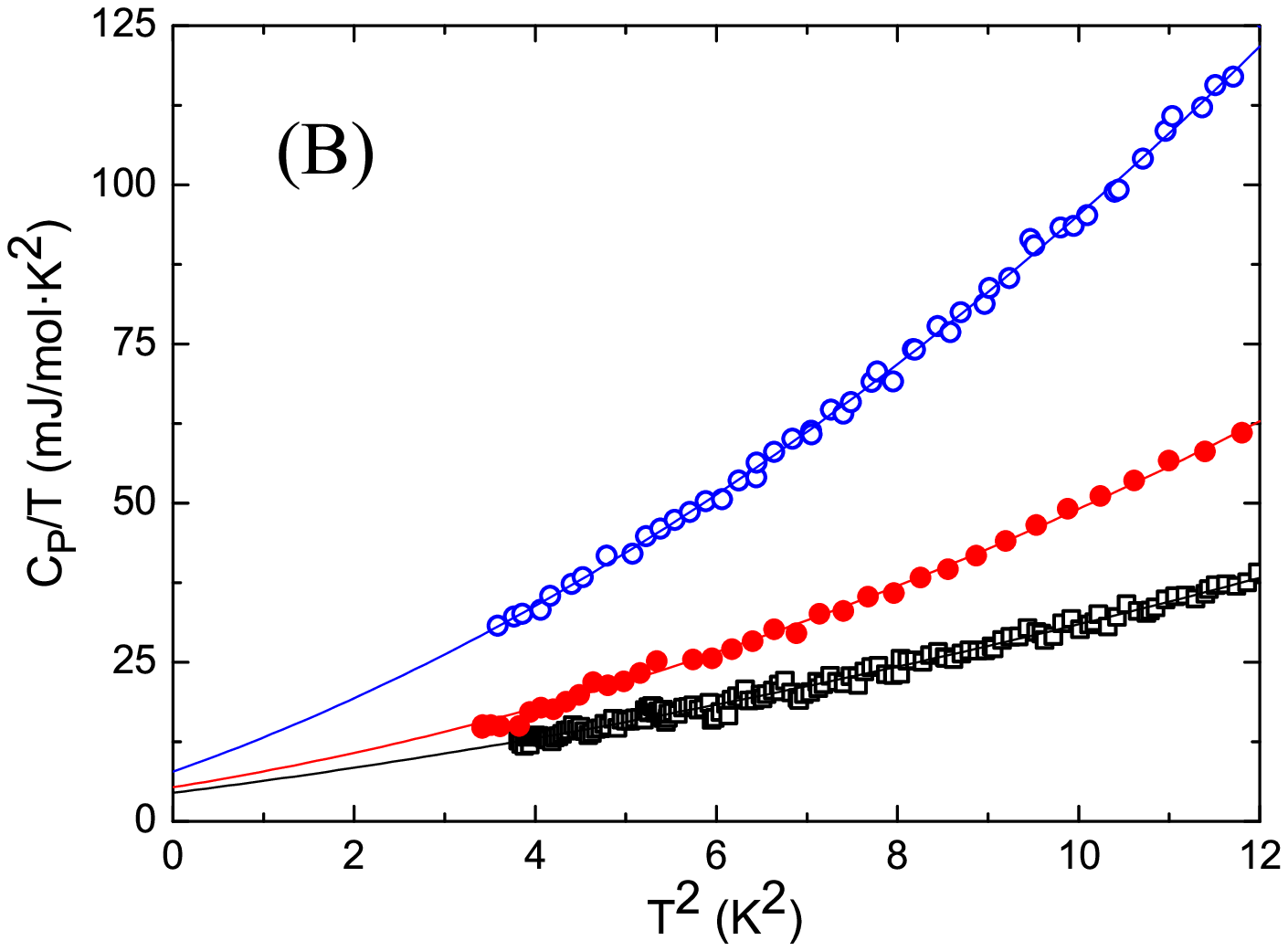}
		 
         \caption{(A) Debye-reduced specific heat $C_p/T^3$ for the monoclinic crystals of CBr$_2$Cl$_2$ (open blue circles), CBrCl$_3$ (solid red circles) and CCl$_4$ (open black squares). The peak position is marked by arrows and the corresponding Debye cubic coefficients $C_D$ obtained from the fits in the panel (B) are shown by horizontal dashed lines (same colors). Earlier published data for CCl$_4$ are also shown by solid green stars \cite{atake1971heat}, solid pink stars \cite{Bagatskii1971} and open blue stars \cite{hicks1944heat}. (B) Our specific-heat data, with symbols as in panel (A), but plotted at the lowest temperatures as $C_p/T$ vs $T^2$. The curves have been fitted in the appropriate range to a quadratic polynomial (see text). The obtained coefficients are shown in Table \ref{tab:lista1}.}
\label{fig:Cpexperimental}
\end{figure}

Low-temperature specific-heat data obtained for the three substances CBr$_n$Cl$_{4-n}$ ($n=0,1,2$) in their monoclinic phases are shown in Fig. \ref{fig:Cpexperimental}(A). They are presented as $C_{p}/T^3$ vs $T$ to emphasize the deviation from the expected Debye behavior at low temperature $C_{p} \propto T^3$ valid for insulating stable crystals. The corresponding Debye coefficients $C_D$ (Table 1) are also indicated there by horizontal dashed lines. Earlier published data for CCl$_4$ at not so low temperatures \cite{hicks1944heat,atake1971heat,Bagatskii1971} are also shown for comparison. CBr$_2$Cl$_2$ and CBrCl$_3$ exhibit clear maxima in $C_{p}/T^3$ at about 7.5 and 7.7 K, respectively, that show a clear resemblance with the ubiquitous BP observed in most glasses, including above-mentioned OG or \textit{glassy crystals}. Unexpectedly, CCl$_4$ also exhibits a noticeable and similar peak, though clearly smaller and shifted to 9.2 K. Nonetheless, its size, shape and position are very similar to those found in many glasses and different from the $C_{p}/T^3$ narrower maxima observed at about 12$-$30 K in many crystals \cite{chumakov2011equivalence} due to van Hove singularities \red{(see Figs. 3.3 and 3.10 in Ref. \cite{Phillips1981Amorphous} and Refs. \cite{ramos1997quantitative,talon2002E,talon2002G,hassaine2012low}, as well as Fig. \ref{fig:Cp_all} below)}.

The deviation from the expected Debye behavior of $C_{p}/T^3$ at low temperature of these insulating crystals is emphasized in {Fig. \ref{fig:Cpexperimental}(B)} which plots $C_{p}/T$ vs. $T^2$. A glance for the three materials immediately reveals a significant departure, without any possible resemblance with a crystalline counterpart, which dominates at temperatures below 1$-$2 K \cite{zeller1971thermal,Phillips1981Amorphous}. As said above, the thermal excitations producing this typical glassy feature (and other ones in thermal conductivity or in acoustic and dielectric properties) \cite{Phillips1981Amorphous} are long thought to be \textit{tunneling} TLS \cite{phillips1972tunneling,anderson1972anomalous}, although some criticisms about the precise interpretation of these universal low-energy excitations have been posited long ago \cite{Yu88LowT}.
It has been shown \cite{ramos2004calorimetric} that the best way to assess quantitatively the TLS contribution is to conduct a quadratic fit in the appropriate temperature range below the $C_{p}/T^3$ maximum by using:
\begin{equation}\label{cp}
 C_p = C_{TLS} T + C_D T^3 + C_{sm}T^5 
\end{equation}
where $C_{TLS}$, $C_D$ and $C_{sm}$ coefficients account for the contributions from the TLS, Debye acoustic modes and additional “soft modes”, respectively, following the phenomenological Soft-Potential Model (SPM) \cite{Ramos1998tunneling,ramos2004calorimetric}.

The results from these fits are shown in Table 1, also including the correspondingly obtained Debye temperatures $\theta_D$. It should be stressed that the definition of the Debye temperature $\theta_D$ depends on the number of vibrating particles in the Debye lattice. Although only acoustic phonons should be taken into account to evaluate the \textit{genuine} Debye temperature or frequency of a solid, often in the literature other more loosely ways of counting active phonons are considered, thus providing very different, usually much higher values. A direct comparison between $\theta_D$ values reported by different authors is therefore often misleading. Table \ref{tab:lista1} provides the usual atomic Debye temperatures \red{(see Supplemental Material at [URL will be inserted by publisher] for discussion of different approaches of Debye temperature)\cite{ashcroft1978solid,kittel1996introduction,hassaine2012low,ohta1995heat,krivchikov2015effects}.}
%, while \textit{Supp Mat} discusses the different approaches}.
%experimental Cp

%%%%%%%%%

%table coefficient
\begin{table}[h!]

\centering
\caption{$C_{TLS}$, $C_D$ and $C_{sm}$ are the linear, cubic and fifth-power coefficients, respectively, from a quadratic fit of $C_p/T$ vs $T^2$ at low enough temperatures (see text for details). $T_{max}$ is the temperature at which $C_p/T^3$ exhibits a broad maximum and $\Theta_D$ is the Debye temperature directly obtained from the cubic coefficient $C_D$ of Eqn. \ref{cp}.}

\begin{tabular}{|c|c|c|c|c|c|}

\hline
\hline 
	& $C_{TLS}$ & $C_D$ & $C_{sm}$ & $T_{max}$ & $\Theta_D$\\
    & (mJ/mol$\cdot$ K${}^2$) & (mJ/mol$\cdot$ K${}^4$) & (mJ/mol$\cdot$K${}^6$)& (K) & (K)\\
\hline
CCl$_4$&4.5 $\pm$ 0.4&1.81 $\pm$ 0.09&  0.084 $\pm$ 0.004&9.2& 175 \\
CBrCl$_3$&5.4 $\pm$ 1.2&2.3 $\pm$ 0.3&  0.21 $\pm$ 0.02&7.7& 162 \\ 
CBr$_2$Cl$_2$&7.8 $\pm$ 1.6&5.0 $\pm$ 0.4&0.37 $\pm$ 0.03&7.5& 125 \\ 
\hline 
\end{tabular}
	\label{tab:lista1}
%\vspace{0.1cm}
\hspace{-0.4cm}
\end{table}
%%%%%%

The data listed in Table \ref{tab:lista1} show that all the three compounds display relatively large values for the $C_{TLS}$ coefficient if compared with the Debye term. The ratio $C_D/C_{TLS}$ increases from 0.4 to 0.64 as the molecular mass is increased and the molecule point group symmetry is lowered. In contrast $C_D/C_{sm}$ shows a decreasing value with increasing molecular mass.

Data displayed in Fig. \ref{fig:Cp_all} emphasizes the deviation from harmonic, Debye behavior as manifested by an upturn of the measured quantity for temperatures below the minimum (corresponding to temperatures below 3 K), an upturn without any possible resemblance with a crystalline counterpart (as shown for ethanol and glycerol crystals), as well as the similar shape for the BP of different glasses of different nature.

\begin{figure}[h!]
	\centering
    %Cp/T3  
		 \includegraphics[width =0.95\columnwidth , clip = true, trim = 0.1cm 0cm 0cm 0cm]{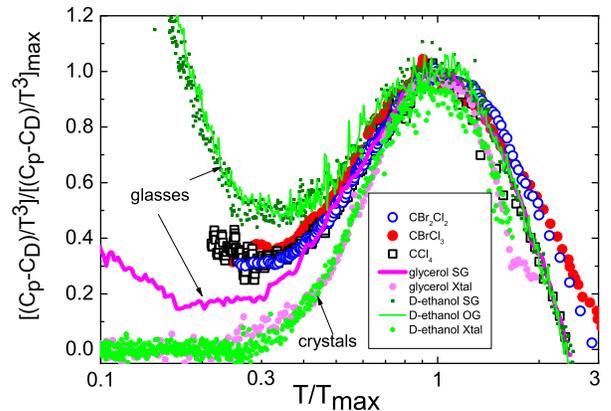}
    		 
 \caption{Normalized specific-heat data referred to $C_D$ values, scaled to the height $[(C_p-C_D)/T^3)]/[(C_p-C_D)/T^3)]_{max}$ and shifted to the position of the peak to show the universal shape for the three halomethanes, CCl$_4$ (black empty squares), CBrCl$_3$ (full red circles) and CBr$_2$Cl$_2$ (blue empty circles), for several structural glasses as glycerol\cite{talon2002G} (pink line) and deuterated ethanol (green dots), orientational glass of deuterated ethanol\cite{talon2002E} (continuous green line) and several ordered crystalline phases of glycerol\cite{talon2002G} (pink circles) and ethanol (green circles)\cite{talon2002E}.}
\label{fig:Cp_all}
\end{figure}
%Normalized specific-heat data referred to $C_D$ values, scaled to the height $[(C_p-C_D)/T^3)]/[(C_p-C_D)/T^3)]_{max}$ and shifted to the position of the peak to show the universal shape for the three halomethanes, CCl$_4$ (black empty squares), CBrCl$_3$ (full red circles) and CBr$_2$Cl$_2$ (blue empty circles), for several structural glasses as glycerol (pink line) and deuterated ethanol (continuous green line), orientational glass of deuterated ethanol (dotted green line) and several ordered crystalline phases of glycerol (pink circles) and ethanol (green circles).

Inelastic neutron scattering (INS) measurements were undertaken by means of the TOSCA and MARI spectrometers at ISIS (details can be found above).

\begin{figure}[h!]
	\centering
    %TOSCA
		 \includegraphics[width = \columnwidth , clip = true, trim = 0cm 0cm 0cm 0cm]{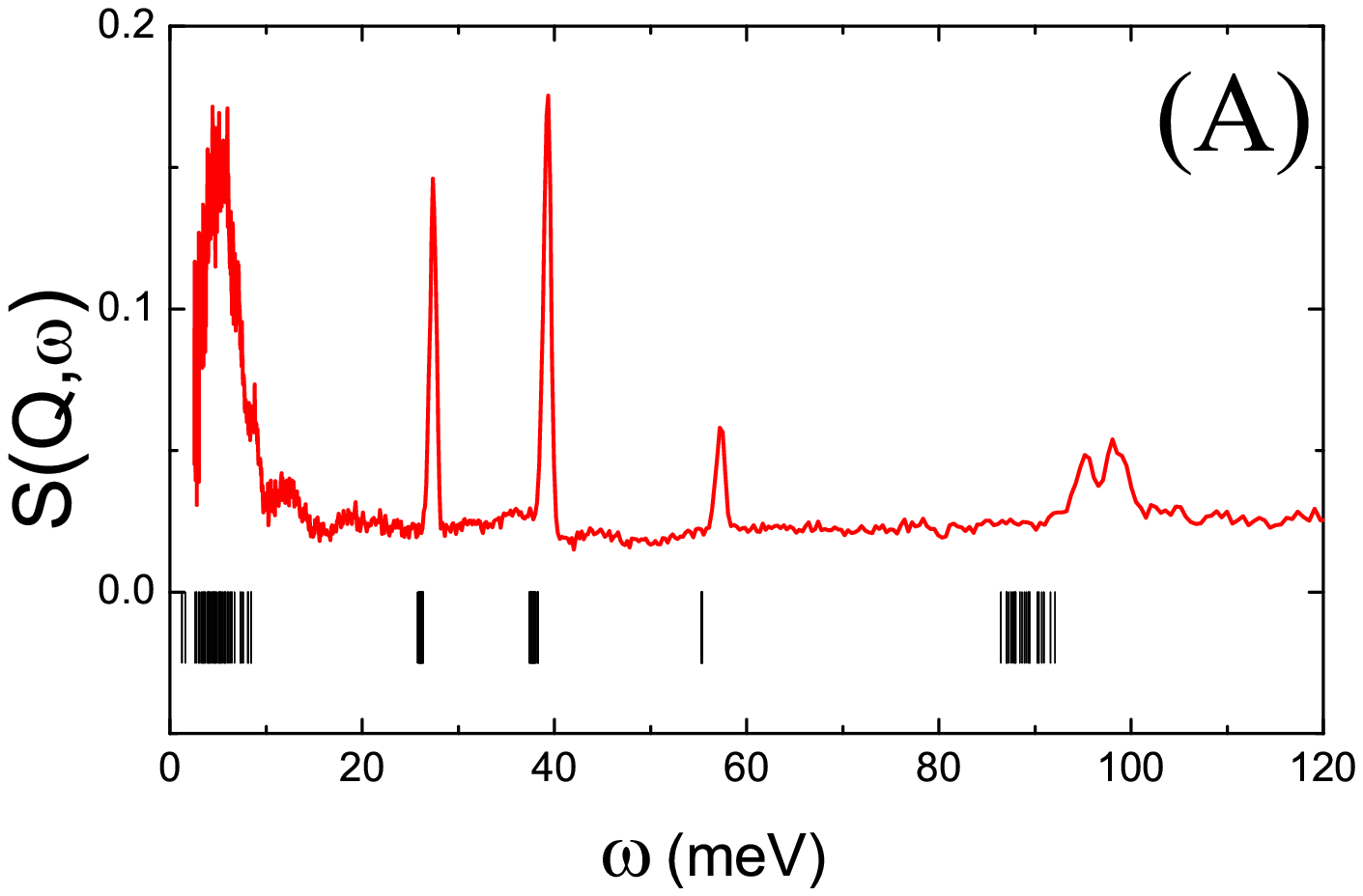}
    %g(w) MARI
		 \includegraphics[width = \columnwidth , clip = true, trim = 0cm 0cm 0cm 0cm]{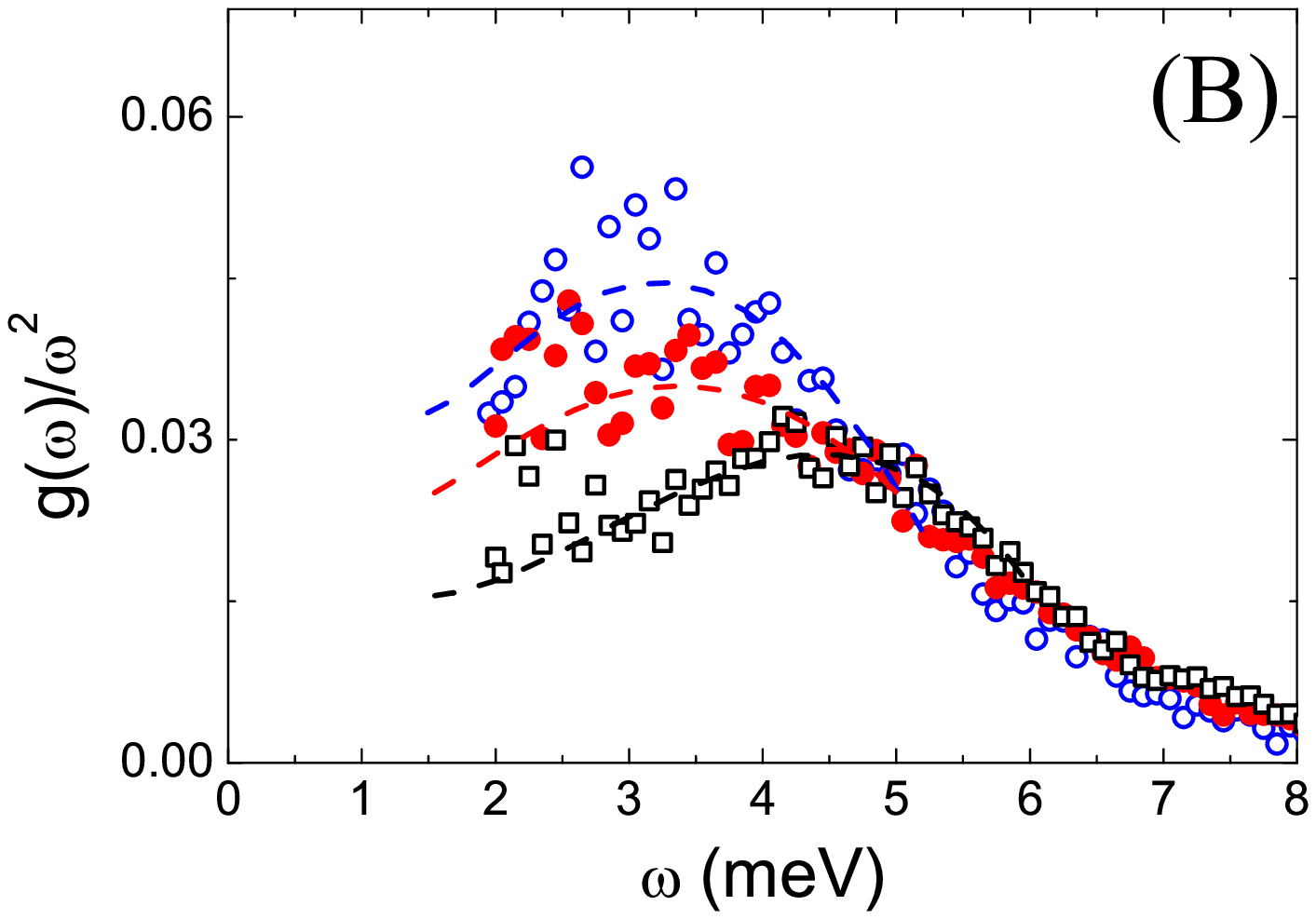}
	
         \caption{(A) Vibrational spectrum as measured using TOSCA spectrometer at 10 K (red) with the calculated quantity at the crystal $\Gamma$-point (black sticks) (see text). (B) Low-frequency vibrational spectra as measured using MARI spectrometer in the reduced form $g(\omega)/\omega^2$, for the monoclinic phases of CCl$_4$ (black empty squares), CBrCl$_3$ (full red circles) and CBr$_2$Cl$_2$ (blue empty circles). Dashed lines are a guide to the eye.}
\label{fig:TOSCA&MARI}
\end{figure}

Fig. \ref{fig:TOSCA&MARI}(A) displays a comparison of the $g(\omega)$ frequency distribution derived from the DFT calculation with the $S(Q,\omega)$ spectrum measured using the TOSCA spectrometer at 10 K for CCl$_{4}$ \cite{DYMKOWSKI2018}. The spectrum extends up to 124 meV although all features above some 12 meV contain contributions from internal molecular high-frequency modes which are beyond the scope of the present study. The spectral region of interest thus comprises frequencies below some 10 meV. The calculated spectrum comprises purely harmonic, one-phonon, single scattering events only, and is not affected by sizeable resolution or attenuation (i.e. Debye-Waller) effects as it is the case for the experimental data. Such effects account for most of the significant differences between experiment and calculation. An improved measurement of the low-frequency part of the distribution is provided in Fig. \ref{fig:TOSCA&MARI}B which compares the results with measurements carried out using a chopper spectrometer (MARI). The region below 10 meV should therefore comprise all the relevant features as far as glassy dynamics is concerned.

The Debye-reduced $g(\omega)/\omega^2$ in Fig. \ref{fig:TOSCA&MARI}B shows peaks at frequencies $\omega_{max}$ around 4.3, 3.3 and 3.1 meV for $n=0,1,2$, respectively. These peaks in the VDoS are in very good agreement with the corresponding peaks at $T_{max}$ in $C_p/T^3$, since they follow the expected numerical factor around 4$-$5 between them, i.e., $\hbar\omega_{max} = (4-5)k_{B}T_{max}$, as occurs for the BPs in glasses \cite{chumakov2009contribution,carini2016low}.

\begin{figure}[h!]
	\centering
    %Dispersion+vDOS
		 \includegraphics[width = \columnwidth , clip = true, trim = 0cm 0cm 0cm 0cm]{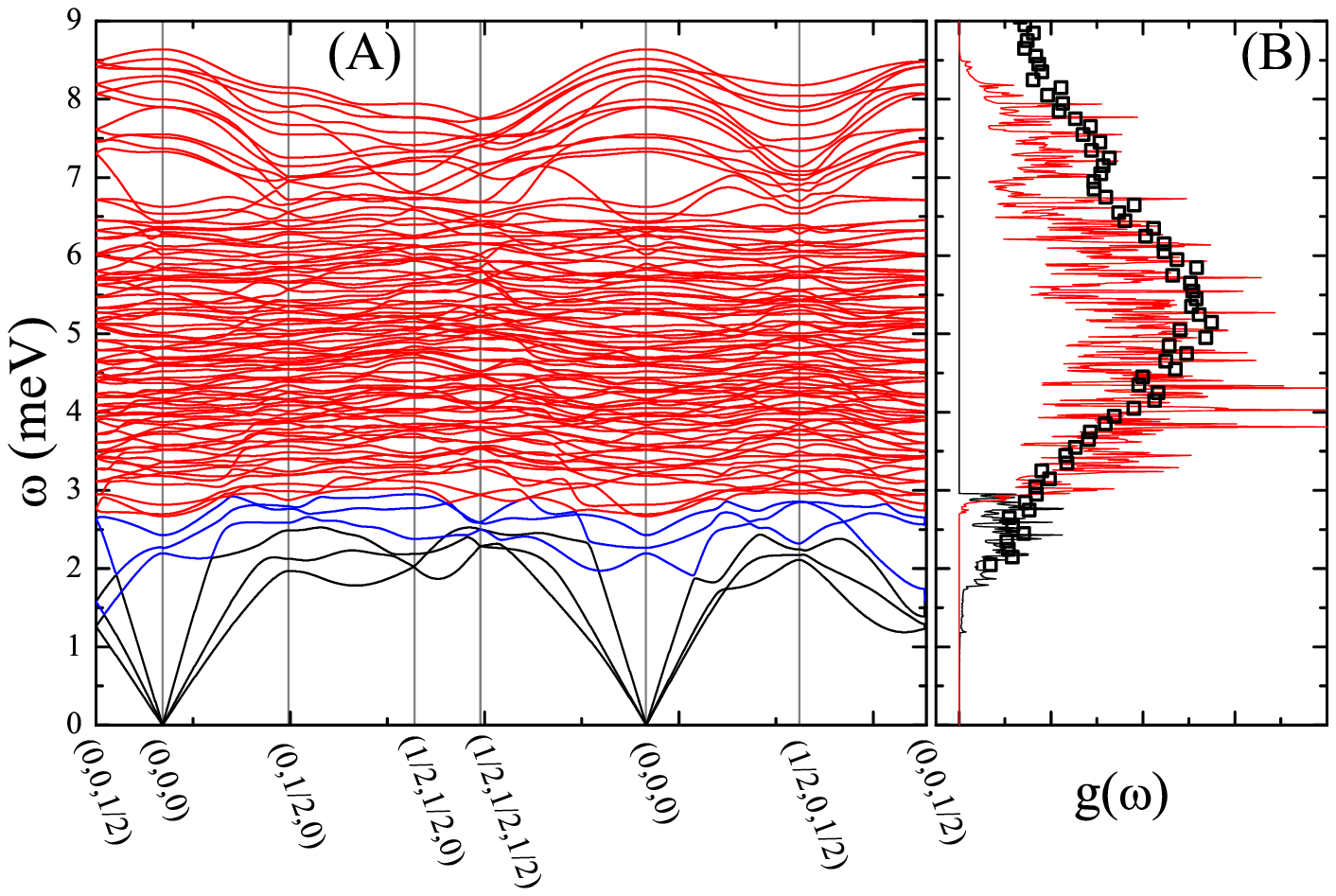}
    %g(w) DISPERSIONS up to 3meV
		 %\includegraphics[width = \columnwidth , clip = true, trim = 0cm 0cm 0cm 0cm]{Disp_3meV_isotops.png}
		\includegraphics[width = \columnwidth , clip = true, trim = 0cm 0cm 0cm 0cm]{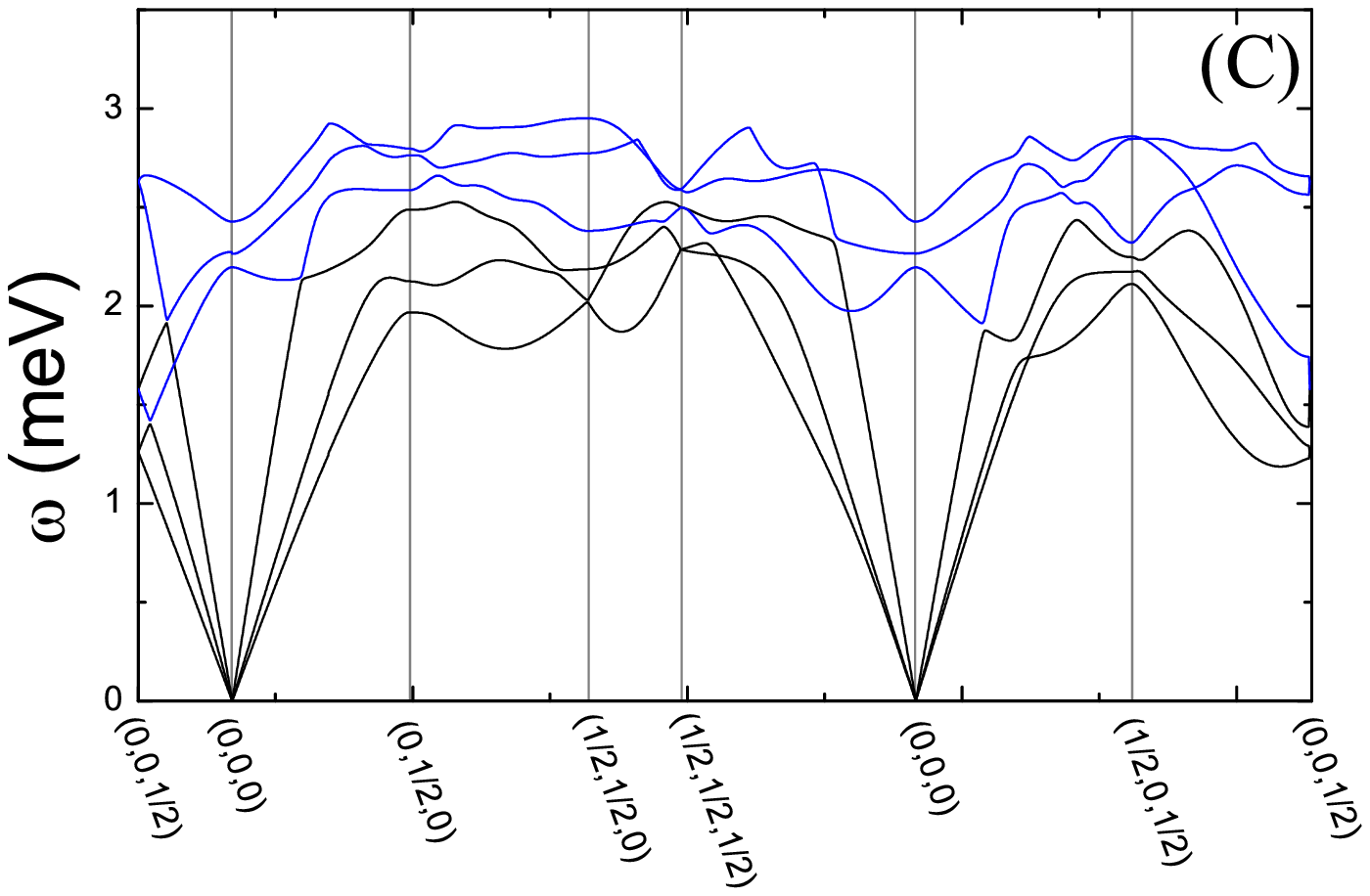}
         \caption{(A) Dispersion relations for C${}^{35}$Cl$_{3}{}^{37}$Cl obtained from DFT. (B) Vibrational density of states calculated from the dispersion relations (black for the three acoustic branches plus the first three optical branches, and red for the contributions of all branches) and experimental as measured from inelastic neutron scattering using MARI spectrometer (black squares). (C) Relation dispersions in the low-energy range for the acoustic branches (black) and the first three optical branches (blue).}
\label{fig:Dispersions}
\end{figure}

The \textit{ab-initio} lattice dynamics calculations using DFT for the reference case of CCl$_{4}$ are shown in Fig. \ref{fig:Dispersions}. In the first Brillouin zone, in addition to the ubiquitous acoustic phonon branches, there are 477 optical phonons (see Figs. \ref{fig:Dispersions}A and \ref{fig:Dispersions}C). The most striking feature concerning Fig. \ref{fig:Dispersions} regards the large number of branches lying between $\approx2$ meV and $\approx9$ meV as well as the manifold of excitations which appear in the calculation as a broad band centred at about 89 meV (not shown) which corresponds to a resolved multiplet in the experimental spectra centered at some 96 meV. It arises from the triply-degenerated $\nu_3$ band shown by the free molecule, comprising three out of the nine internal molecular modes which represent motions where the molecular center of mass gets displaced from its equilibrium position. The number of branches within 3$-$9 meV contrasts with the expectancy of a far more reduced number of motions of librational or librational-translational character expected for known, fully ordered molecular crystals \cite{Decius1977Molecular}, although it seems to be a common feature for materials of lower dimensionality \cite{APMayer2004JPhys}.
The natural compound C${}^{35}$Cl$_{3}{}^{37}$Cl, with C$_{3v}$ molecular point symmetry and with an exchange between sites occupied by ${}^{35}$Cl and ${}^{37}$Cl, that makes it similar to the CBrCl$_3$ case, qualitatively displays the same dispersion relations that pure C${}^{35}$Cl$_{4}$, giving rise only to an overall lowering of the energy (softening) of acoustic and optical modes \red{(see Supplemental Material at [URL will be inserted by publisher])}, but with a non-essential effect in the VDoS behavior. It can also be noticed (see Fig. \ref{fig:Dispersions}A) that the low-energy (\textit{soft}) modes at about 3$-$4 meV, responsible for the BP-like, mainly arise from the lowest optical modes that merged with Brillouin-zone-boundary acoustic phonons. The observed “piling up” of optical modes makes it impossible to rationalize the BP-like feature in terms of a modification of the van Hove singularity as some precedent works proposed \cite{chumakov2011equivalence}. 

The characteristic frequencies at $Q=0$ for the set of the thirteen lowest energy branches are bounded below 3.55 meV, and comprise a group of ten optical branches with energies down to 2.33 meV. To deepen into the character of such low-frequency modes, the mode eigenvectors resulting from the solution of the eigenvalue problem have been analyzed in some detail. The result of such an exercise is best gauged by means of direct visualization of the time dependence of molecular motions within the crystal lattice as given by a set of motion pictures \red{(see Supplemental Material at [URL will be
 inserted by publisher] for movies)}. A glance to such graphical representations reveals the following. First, as it could be expected, all such modes show a mixed rotation-translation character also including minor molecular deformation contributions. Furthermore, the reorientational motions do involve significant displacements of the molecular centers of mass, a finding which is glaringly exemplified by the complete lifting of the degeneracy of the higher-frequency $\nu_3$ band of individual molecules. Such results vividly exemplify how strong translational-rotational coupling dominates molecular motions with a crystal lattice even for considerably large frequencies. A result which merits to delve attention at, concerns the three lowest frequency motions, although such features are also found at higher frequencies. The analysis of the mode eigenvectors for such motions has revealed that the angular excursions of single molecules are coupled to highly cooperative out-of-phase displacements of pairs of crystal planes. Finally, the highly cooperative character of the set of optical modes here considered is progressively lost as the mode energy increases while molecular deformational components increase in importance. The result may appear somewhat surprising since such modes have energies some 22 meV below that of the lowest-lying normal mode of an isolated molecule.

\begin{figure}[h!]
	\centering
    %DFT Cp
		 \includegraphics[width = \columnwidth , clip = true, trim = 0cm 0cm 0cm 0cm]{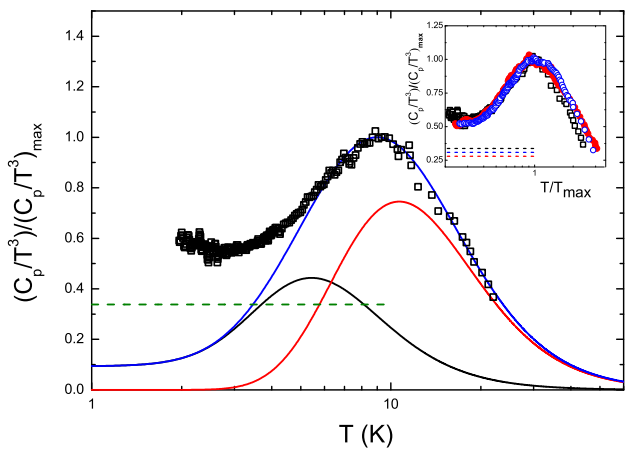}
		 
         \caption{Experimental specific-heat data (black squares) at the lowest temperatures plotted as normalized $(C_p/T^3)/(C_p/T^3)_{max}$ vs. $T$ for the monoclinic phase of CCl$_4$. Solid lines correspond to DFT calculated contributions for the reference case of C${}^{35}$Cl$_{3}{}^{37}$Cl: black curves shows the contribution of the acoustic plus the three low-energy optical branches, red curve shows the contributions of the rest of optical branches and blue curve correspond to the total (acoustic and optical) contributions of the phonons to the specific heat. Horizontal dotted line corresponds to the normalized Debye value. Inset: Experimental specific-heat data at the lowest temperatures plotted as normalized $(C_p/T^3)/(C_p/T^3)_{max}$ vs. $T/T_{max}$ for CCl$_4$ (black squares), CBrCl$_3$ (red circles) and CBr$_2$Cl$_2$ (blue circles).}
\label{fig:Cp_calc}
\end{figure}

In Fig. \ref{fig:Cp_calc}, the Debye-reduced specific-heat data of crystalline C${}^{35}$Cl$_{3}{}^{37}$Cl, $C_p/T^3$, scaled to its height value, are compared to the calculated values by DFT, where the contribution of acoustic phonons and optical phonons to the specific heat is depicted separately, besides showing the total.

It is important to stress that the excellent agreement with experimental data of our DFT calculations for the $C_p/T^3$ BP-like shape and position (Fig. \ref{fig:Cp_calc}) is not the result of any fit nor any temperature scaling. Moreover, the consistency of both experimental INS data and DFT calculations is also confirmed here by their mutual agreement in the resulting VDoS, $g(\omega)$, as shown in Fig. \ref{fig:Dispersions}B. In particular, DFT results enables to separate the contribution of the lowest frequency modes (three acoustic and the three lowest lying optical branches) from the manifold of modes above 3 meV to the specific heat, revealing that the dynamics of a perfectly ordered, harmonic crystal accounts for the measured specific heat down to some 6 K.
	Finally, it could be remarked that we are finding essentially the same set of glassy features ubiquitously observed in non-crystalline solids of disputed origin. Nonetheless, in our case the low-frequency vibrational modes producing the BP-like feature are identified as genuine low-lying optical phonons within a genuine Brillouin zone merged with near-boundary acoustic phonons, producing equivalent effects. The current results are thus of relevance for current discussions on the dynamics of glassy matter. The comparison of the experimental data for $C_p$ with those calculated under the harmonic approximation using the VDoS derived from the lattice dynamics results show that features appearing in plots of $C_p/T^3$ such as well defined maxima, sometimes dubbed as "boson peaks", arise in the current case, from the dense manifold of optic-like branches above some 2$-$3 meV for CCl$_4$ as well as for the occupationally disordered monoclinic phases of CBrCl$_3$, and CBr$_2$Cl$_2$. 
The access to detailed microscopic information provided by the DFT calculations concerning rather low-frequency optic modes revealed a strikingly strong coupling of molecular reorientations with out-of-phase displacements of crystal planes. We would expect that some remnants of such motions will be present in the real, dynamically disordered crystal and thus provide a strong coupling mechanism of molecular reorientations to low-frequency acoustic phonons. Such finding may thus be of relevance for ascribing a microscopic origin to motions giving rise to the observed anomalies in the specific heat.

\red{Finally, it should be mentioned that the (harmonic) DFT lattice dynamics calculations cannot describe the upturn of the $C_p/T^3$ at low-temperature (below 2 K) corresponding to tunneling TLS. This upturn, experimentally evidenced for the three title compounds, is also a glassy fingerprint of the here presented crystals.}

\section{\label{sec:Conc}Conclusion}

In summary, the major finding of our work is that the three studied CBr$_n$Cl$_{4-n}$ monoclinic crystals exhibit the characteristic glassy behavior at low temperature, namely a broad (“boson”) peak in both $g(\omega)/\omega^2$ and $C_p/T^3$, and a non-zero linear term $C_{TLS}$ well beyond experimental error, which increase with the number $n$ of Br atoms per molecule, $n = 0,1,2$. For CBrCl$_3$ and CBr$_2$Cl$_2$, a random occupancy of the halogen atoms appears to distort the dynamical network as to produce these low-frequency excitations. The freezing-in of this atomic disorder appears to be the origin of the glassy signatures as those emerging in canonical, structural glasses. 
	Even more surprisingly, the subtle dynamical disorder involving the chlorine atoms in the “well-ordered” low-temperature phase of CCl$_4$ is able to produce the same glassy phenomenology, though to a lesser extent. Being CCl$_4$ a fully-ordered crystal, we have been able to determine and study the dispersion relations for acoustic and optical phonons over the whole Brillouin zone. The picture that emerges portrays molecular motions within this material as able to couple to the acoustic field up to surprisingly large frequencies. Such coupling leads to a lift of the degeneracy of individual crystal and molecular modes giving rise to a dense mesh of optic branches at low and even relatively high frequencies. The former are shown to give rise to the peak in $C_p/T^3$, as well to a peak in the frequency distribution $g(\omega)/\omega^2$. Although this conclusion is applicable \textit{stricto sensu} only to the studied CCl$_4$ crystal, the universality found in the \textit{glassy anomalies} makes it very likely that some lessons or hints can be drawn from these findings to the much debated issue of the boson peak and other glassy anomalies in structural glasses and disordered crystals. Moreover, our results are at variance with some of the recent proposals \cite{chumakov2011equivalence} which suggest the attribution of glassy features to van Hove-like singularities for acoustic phonons.
	
\section*{Acknowledgments}

The authors are grateful for the financial support received within the projects FIS2017-82625-P and FIS2017-84330-R from MINECO. We also acknowledge the Generalitat de Catalunya under Project 2017SGR-042 and the Autonomous Community of Madrid through program NANOFRONTMAG-CM (S2013/MIT-2850). One of us (M.A.R.) also acknowledges the “María de Maeztu” Program for Units of Excellence in R\&D (MDM-2014-0377).

The authors gratefully acknowledge the {\it Science \& Technology Facilities Council} for financial support for this project, including personnel training and visits to the ISIS Molecular Spectroscopy Group, access to beam time at the ISIS Facility (RB910454, RB1320105), as well as computing resources from the STFC Scientific Computing Department's SCARF cluster.

\bibliography{Halomethane_PRB}
\end{document}